\documentclass[reprint,superscriptaddress,amsmath,amssymb,aps,showpacs,prb]{revtex4-1}

\usepackage{graphicx}
\usepackage{bm}
\usepackage{epsfig}
\usepackage{amssymb}
\usepackage{amsfonts}
\usepackage{braket}
\usepackage{color}
\usepackage{natbib}
\usepackage[caption=false]{subfig}

\newcommand{\vep}{\varepsilon}
\newcommand{\la}{\langle}
\newcommand{\ra}{\rangle}
\newcommand{\nn}{\nonumber}

\graphicspath{{figures/}}% Put all figures in this directory. Avoids clutter.

\begin{document}

\date{\today}
\title{Magnetic Quantum Phase Diagram of Magnetic Impurities in 2
  Dimensional Disordered Electron Systems}
\author{Hyun Yong Lee}
\affiliation{Division of Advanced Materials Science, Pohang University
 of Science and Technology (POSTECH), Pohang 790-784, South Korea}
\email[]{hyunyongrhee@postech.edu}
\author{Stefan Kettemann}
\affiliation{Division of Advanced Materials Science, Pohang University
 of Science and Technology (POSTECH), Pohang 790-784, South Korea}
\affiliation{School of Engineering and Science, Jacobs University
 Bremen, Bremen 28759, Germany} 
\email[]{s.kettemann@jacobs-university.de}

\begin{abstract} 
  The quantum phase diagram of disordered electron systems as function of the
  concentration of magnetic impurities $n_m$ and the local exchange coupling $J$
  is studied in the dilute limit. We  take into account the Anderson
  localisation of the electrons by a nonperturbative numerical treatment of the
  disorder potential. The competition between RKKY interaction $J_{\rm RKKY}$
  and the Kondo effect, as governed by the temperature scale $T_K$, is known to
  gives rise to a rich magnetic quantum phase diagram, the Doniach diagram. 
  Our numerical calculations show that in a disordered system both the Kondo
  temperature $T_K$ and $J_{\rm RKKY}$ are widely distributed.
  Accordingly, also their ratio, $J_{\rm RKKY}/T_K$ is widely distributed as
  shown in Fig.\,\ref{fig:ratio_kondo_rkky_square}\,(a). 
  However, we  find a sharp cutoff of
  that distribution, which allows us to define a critical density of magnetic
  impurities $n_c$ below which Kondo screening wins at all sites of the system
  above a critical coupling $J_c$, forming the Kondo phase\,[see
  Fig.\,\ref{fig:ratio_kondo_rkky_square}\,(b)]. As disorder is increased, $J_c$
  increases and a spin coupled phase is found to grow at the expense of the
  Kondo phase. From these distribution functions we derive the magnetic
  susceptibility which show anomalous power law behavior. In the Kondo phase
  that power is determined by the wide distribution of the Kondo temperature,
  while in the spin coupled phase it is governed by the distribution of $J_{\rm
    RKKY}$.  At low densities and small $J< J_c$ we identify a paramagnetic phase. 
  %At higher concentrations $n_m$ a transition to a spin glass phase, followed by
  %a transition to a magnetic phase with long range order is expected. 
  We also report results on a honeycomb lattice, graphene, where we find
  that the spin coupled phase is more stable against Kondo screening, but is more
  easily destroyed by disorder into a PM phase. 
\end{abstract}

\maketitle

\section{Introduction}
Phenomena which emerge from the interplay of strong correlations and disorder
remain a challenge for condensed matter theory. Spin correlations and disorder
effects are however relevant for a wide range of materials, including doped
semiconductors like Si:P close to metal-insulator transitions,\cite{Loehneysen}
and heavy Fermion systems, materials with 4f or 5f atoms, notably Ce, Yb, or
U.\cite{Vojta} Many of these materials show a remarkable magnetic
quantum phase transition which can be understood by the competition between
indirect exchange interaction, the Ruderman-Kittel-Kasuya-Yoshida\,(RKKY)
interaction between localised magnetic moments\cite{Kittel, Kasuya, Yoshida} and
their Kondo screening. 
\begin{figure}[!Hb] 
 \captionsetup[subfloat]{font = {bf,up}, position = top,
  captionskip=0pt, farskip=0pt} 
 \subfloat[~~~~~~~~~~~~~~~~~~~~~~~~~~~~~~~~~~~~~~~~] 
 {\includegraphics[width=0.33\textwidth]{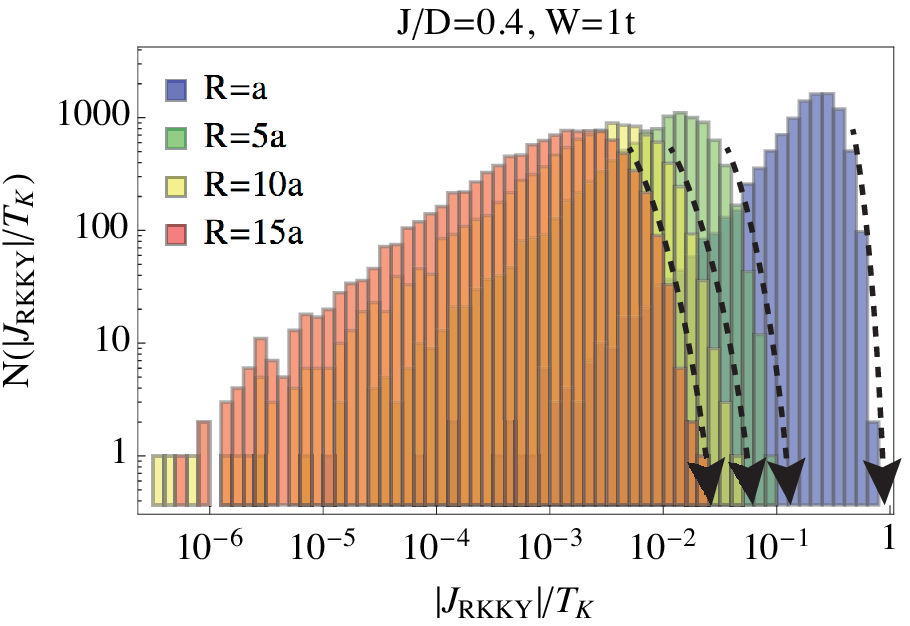}}\\
 \subfloat[~~~~~~~~~~~~~~~~~~~~~~~~~~~~~~~~~~~~~~~~] 
 {\includegraphics[width=0.33\textwidth]{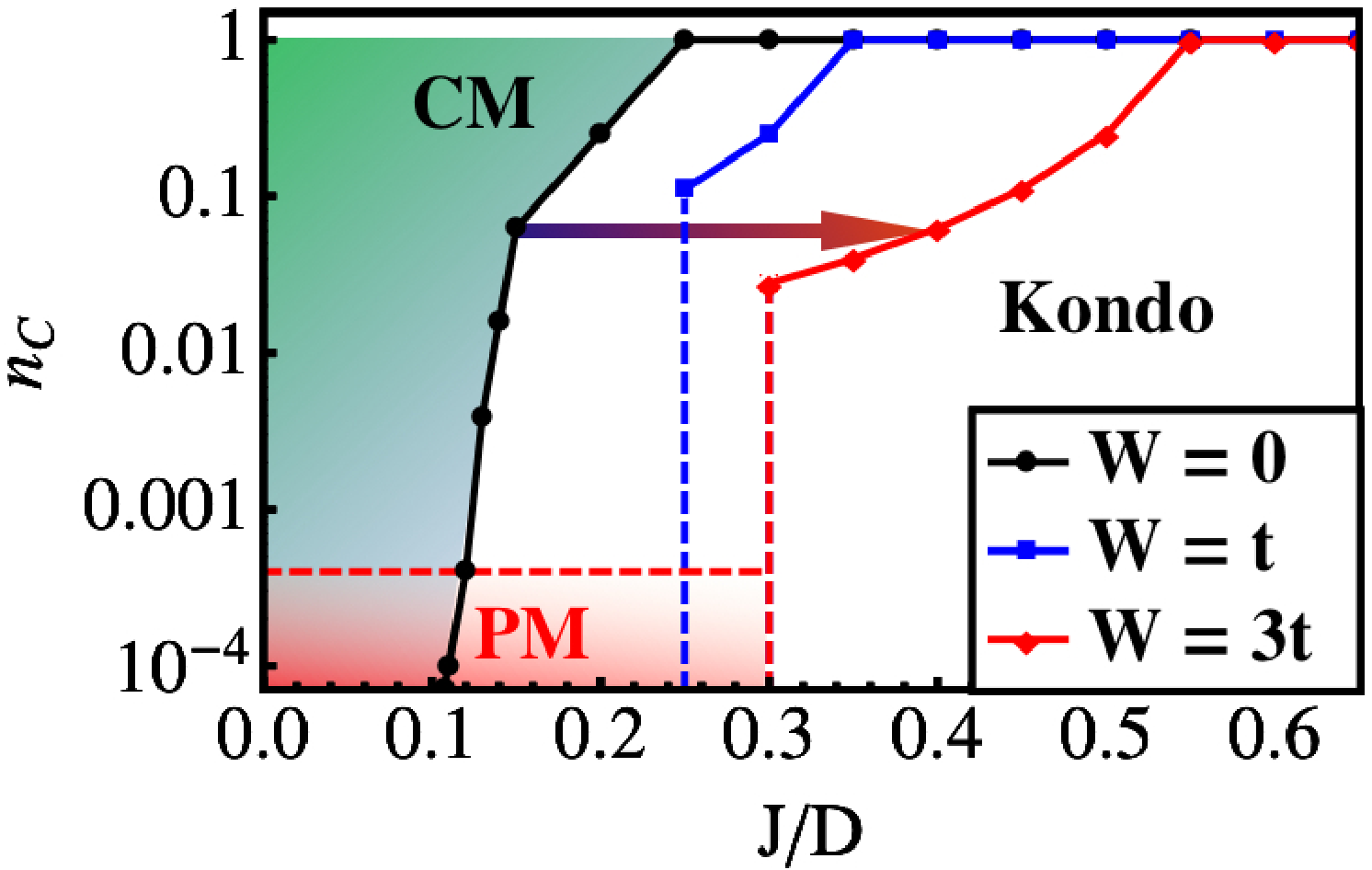}}
 \caption{(Color online) (a) Distribution of the ratio between $|J_{\rm
     RKKY}|$ and $T_K$ for various distances R. Black dashed arrow:
   sharp cutoff of each distribution (b) Magnetic quantum
   phase diagram: critical MI density $n_c$ as function of $J/D$ for
   various disorder strengths $W$ as determined by the distance $R$
   below which $|J_{\rm RKKY}|/T_K$ does not exceed 1 at any
   site. Horizontal dashed line: density $n_{\xi}$ below which there
   is a paramagnetic moment phase (PM). CM= phase with coupled
   magnetic moments.}  
 \label{fig:ratio_kondo_rkky_square}
\end{figure}
Thereby, one finds a suppression of long range magnetic
order when exchange coupling $J$ is increased and Kondo screening succeeds. This
results in a typical quantum phase diagram with a quantum critical point where
the $T_c$ of the magnetic phase is vanishing, the Doniach
diagram.\cite{Doniach} 

Recently, controlled studies of magnetic adatoms on the surface of
metals,\cite{Zhou} on graphene,\cite{Chen} and on the conducting surface of
topological insulators\cite{Hsieh,gap,rader,reinert} with surface sensitive
experimental methods like spin resolved STM and ARPES became possible. This
demands a theoretical study of the Doniach diagram for magnetically doped
disordered electron systems (DES), in particular 2D systems. 
  
In any material there is some degree of disorder. In doped semiconductors it
arises from the random positioning of the dopants themselves, in heavy Fermion
metals and in 2D metals it may arise from structural defects or
impurities. Disorder is known to cause Anderson localisation, which therefore
has to be taken into account when deriving the Doniach diagram of disordered
electrons systems. Moreover, as noted already early,\cite{Anderson} the physics
of random systems is fully described by probability distributions, not just
averages. This must be particularly true for systems with random local magnetic
impurities (MIs),\cite{Mott} since the magnetic impurities are exposed to the
local density of states of the conduction electrons, which is widely distributed
itself. In fact, it has been noticed that a wide distribution of the
Kondo temperature $T_K$ of MIs in disordered host metals gives rise
to non-Fermi liquid behavior, such as the low temperature power-law
divergence of the magnetic
susceptibility.\cite{Mott,Vladimir1,Vladimir2,Bhatt,Wolfle,jones,Grempel,Kettemann1,kskim2}
Nonmagnetic disorder quenches the Kondo screening of MIs due to
Anderson-localisation and the formation of local pseudogaps at the Fermi
energy,\cite{Kettemann1,Kettemann2,Kettemann3,Kettemann5} resulting in bimodal
distributions of $T_K$ and a finite concentration of free, paramagnetic moments
(PMs). However, in these studies the RKKY interaction $J_{\rm RKKY}$ between
different MIs has not yet been taken into account. $J_{\rm RKKY}$ is mediated by
the conduction electrons, and aligns the spins of the MIs ferromagnetically or
antiferromagnetically, depending on their distance $R$. This is a long-ranged
interaction, with a power law decay $J_{\rm  RKKY}\sim 1/R^d$, where d is the
dimension,  and its typical value is not changed by weak disorder.\cite{Lerner,
  Bulaevskii, Bergmann, HYLEE1} However, its amplitude has a wide log-normal
distribution in disordered metals.\cite{Lerner, HYLEE2} In this article we
therefore intend to study the competition between RKKY interaction and the Kondo
effect in disordered electron systems. 

In the next section we introduce the model, and provide the equations for the
Kondo temperature and the RKKY coupling. In section III, we derive numerically
the distribution function of $J_{\rm RKKY}$, and compare it with an analytical result,
based on a perturbative expansion of the nonlinear sigma model. We derive next
numerically the distribution function of $T_K$ finding excellent agreement 
with approximate  analytical results which were obtained, taking into account 
the  multifracatlity and power law correlations of wave functions. In section IV
we present the main results, the  distribution function  of  the Ratio of TK and
RKKY Interaction, for various distances between magnetic impurities R. From that
we show how to derive the zero temperature magnetic quantum phase diagram as
function of magnetic impurity density and exchange coupling, for 2D disordered
electronic systems. At low densities and small $J < J_c$ we identify a
paramagnetic phase. For graphene we find that the spin coupled phase is more
stable against Kondo screening, but is more easily destroyed by disorder into a
PM phase. In section V we derive  from the distribution functions the magnetic
susceptibility as function of temperature, which show anomalous power law
behavior. In the Kondo phase that power is found to be  determined by the wide
distribution of the Kondo temperature, while at small exchange coupling there we 
identify spin coupled phase where the magnetic susceptibility is governed by the
distribution of $J_{\rm RKKY}$. In the final section we conclude and discuss the
relevance and limitations of our results.  

\section{Model}
In order to obtain the Doniach diagram of random electron systems we extend the
approach of Doniach\cite{Doniach} by calculating the distribution functions of 
$T_K$ and $J_{\rm RKKY}$ and their ratio. Thus, in our approach we try to draw
conclusions on the quantum phase diagram of an electron system with a finite
density of magnetic impurities, by considering the Kondo temperature of single
impurities and the RKKY coupling of pairs of magnetic moments. 
       
We start from a microscopic description of the MIs, the Anderson impurity model
coupled to a non-interacting disordered electronic Hamiltonian with on-site
disorder. Then, we map it with the Schrieffer-Wolff transformation on a model of
Kondo impurity spins coupled to the disordered host electron spins by the local
coupling $J$.\cite{Kettemann5} We consider the single-impurity $T_K$ and the coupling 
$J_{\rm RKKY}$ between a pair of spins. For the numerical calculations we employ the
single-band Anderson tight-binding model on a square lattice of size\,$L$ and
lattice spacing\,$a$,  
\begin{equation}
 \centering
 H = -t\sum_{\la i,j \ra} c_{i}^{\dagger} c_{j} + \sum_{i} (w_i - \tilde{E}_F)~c_{i}^{\dagger} c_{i}, 
 \label{eq:Hamiltonian}
\end{equation}
where $t$ is the hopping energy between nearest neighbours $\la i,j \ra$,
% $c_i~(c_i^{\dagger})$ annihilates\,(creates) an
%electron at site $i$, 
$w_i$ is the on-site disorder potential distributed in the interval $[-W/2,
W/2]$. $\tilde{E}_F = E_F + \vep_{\rm edge}$, where $E_F$ is the Fermi
energy measured from the band edge, in 2D $\vep_{\rm edge} = -4t$. We use
periodic boundary conditions.  
%We
% set $\hbar$ and $a$ to unity. 
 
In the dilute limit, one can calculate the $T_K$ of  a single magnetic impurity
at position $\bm{R}_i$ from the Nagaoka-Suhl one-loop equation,\cite{Nagaoka,
  Suhl}  
\begin{equation}
 1 = \frac{J}{2} \int_0^{D} d\vep\, \frac{\tanh[(\vep -
  E_F)/2 T_K]}{\vep - E_F}\,\rho_{ii}(\vep),
 \label{eq:nagaoka_kpm}
\end{equation}
with band width $D$. $\rho_{ii}(\vep) = \braket{i| \delta(\vep - H) |i}$ is the
local density of states\,(LDOS). The RKKY coupling $ J_{{\rm RKKY}_{ij}}$
between two MIs located at positions $\bm{R}_i$,\, $\bm{R}_j$ is in the zero
temperature limit ($T=0$) given by\cite{Didier,HYLEE1} 
\begin{equation}
 J_{{\rm RKKY}_{ij}} = -J^2\frac{S(S+1)}{2S^2} \int_{\vep<E_F} d\vep
 \int_{\vep'>E_F} d\vep' \frac{F(\vep,\vep')_{ij}}{\vep-\vep'}, 
 \label{eq:J-KPM}
\end{equation}
where $F(\vep,\vep')_{ij} = {\rm Re}[\rho_{ij}(\vep)\rho_{ji}(\vep')]$, and $S$
is the magnitude of the MI spin. 

\section{Distribution Functions}

Using the Kernel Polynomial method (KPM),\cite{Didier,Weisse} one can evaluate
the matrix elements of the density matrix $\rho_{ij}(\vep) = \braket{i|\delta
  (\vep-\hat{H})|j}$\cite{HYLEE1, Weisse, Mucciolo} with a polynomial expansion
of order $M$. Here, we increase the cutoff degree $M$ linearly with the linear
system size $L$ based on our analysis for the convergence of RKKY interaction
with respect $M$ in Ref.\,\onlinecite{HYLEE2}. It has been also carefully
discussed in Ref.\,\onlinecite{Wenk} that the choice of $M \propto L$, not $M
\propto L^2$, gives proper DOS and LDOS results avoiding finite size effect.  

\begin{figure}[!Ht]
 \center
 \captionsetup[subfloat]{font = {bf,up}, position = top,
  captionskip=0pt, farskip=0pt} 
 \subfloat[] 
 {\includegraphics[width=0.25\textwidth]{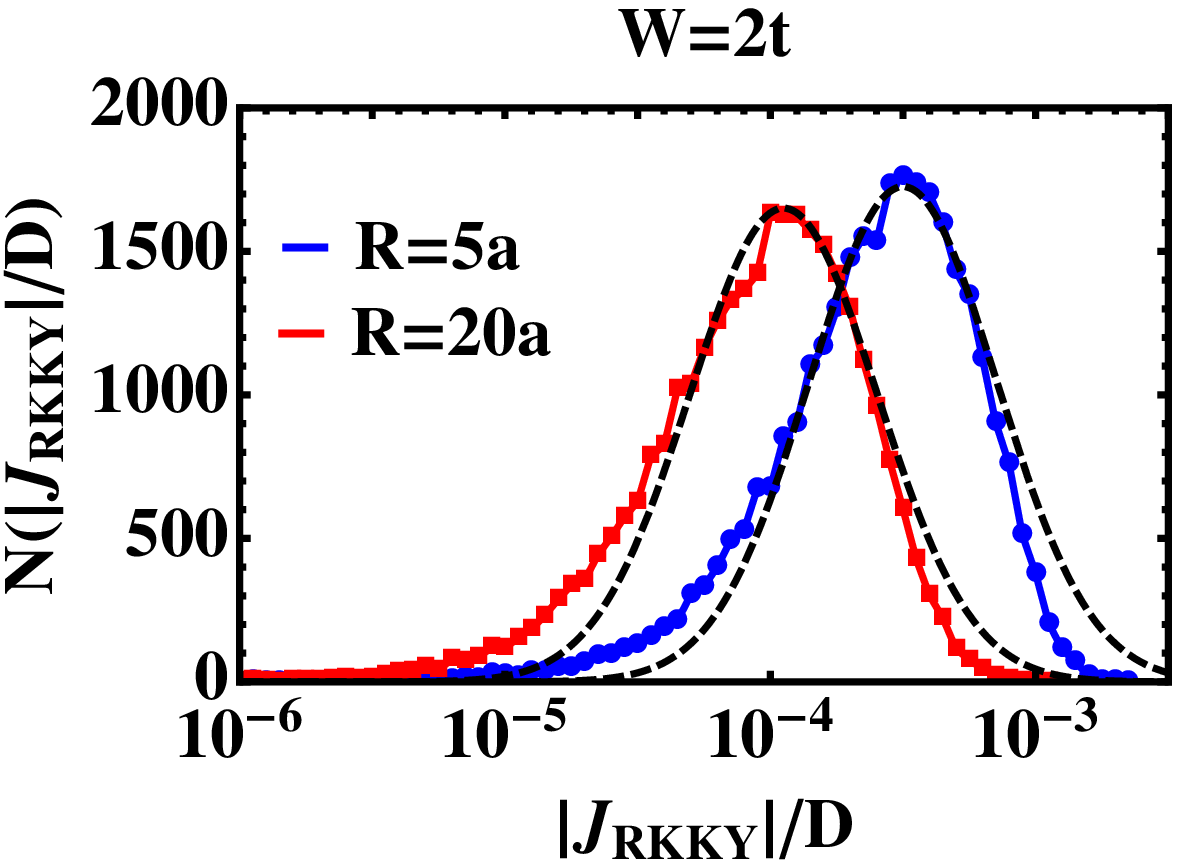}}
 \subfloat[] 
 {\includegraphics[width=0.25\textwidth]{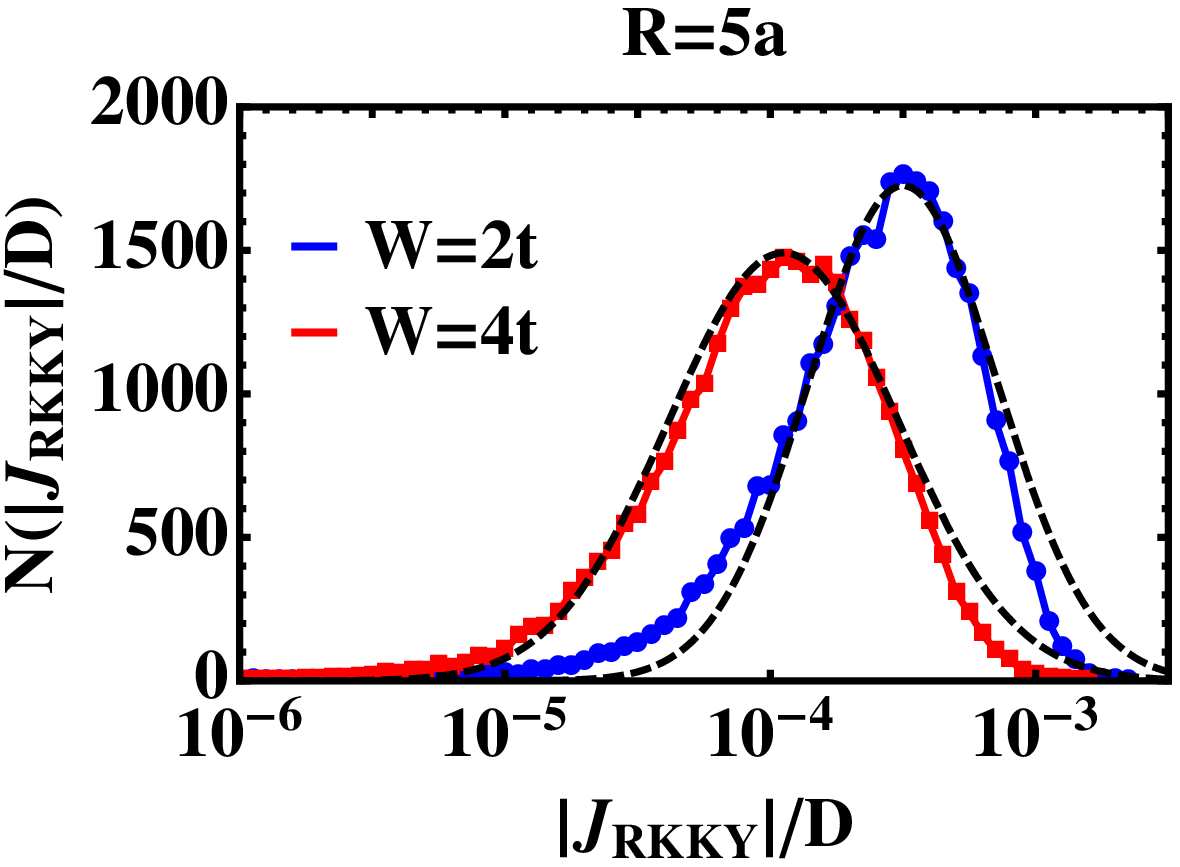}}\\
 \subfloat[] 
 {\includegraphics[width=140px, height=90px]{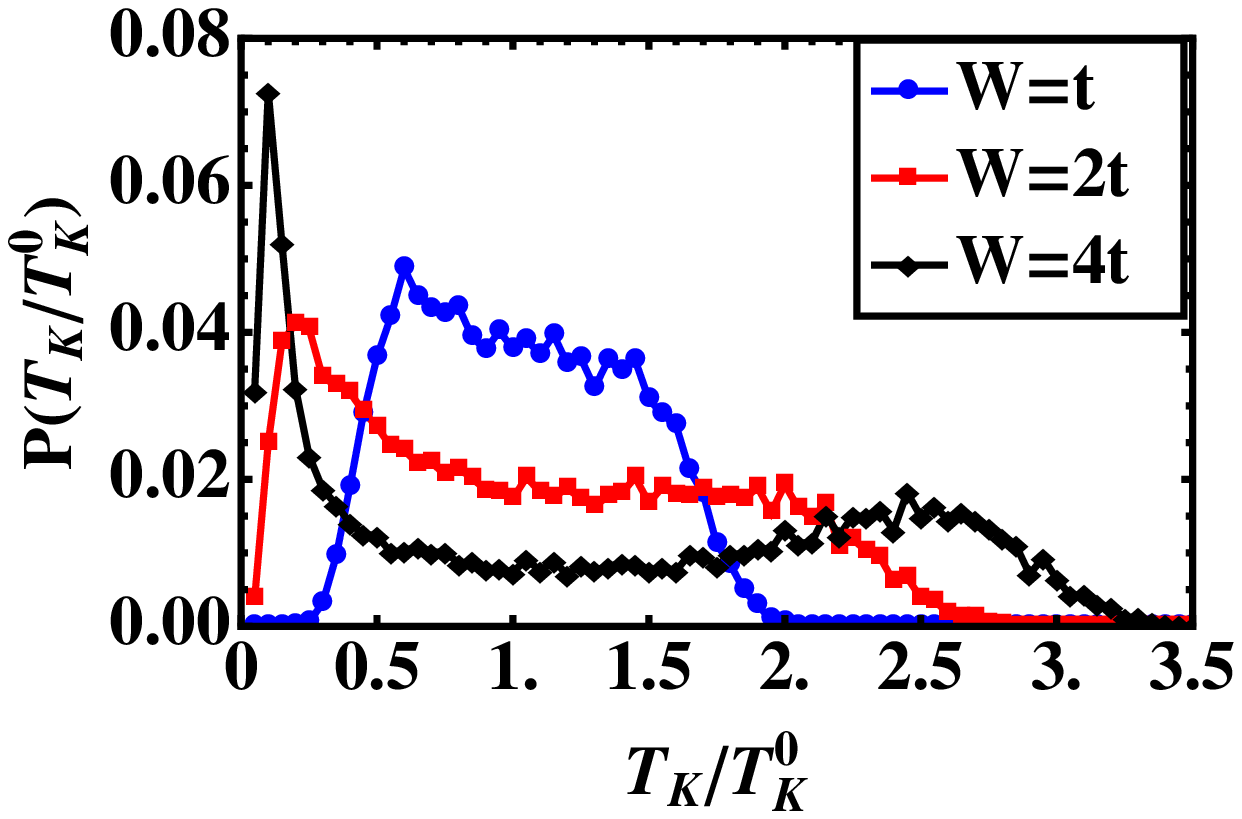}}
 \subfloat[]
 {\includegraphics[width=123px, height=85px]{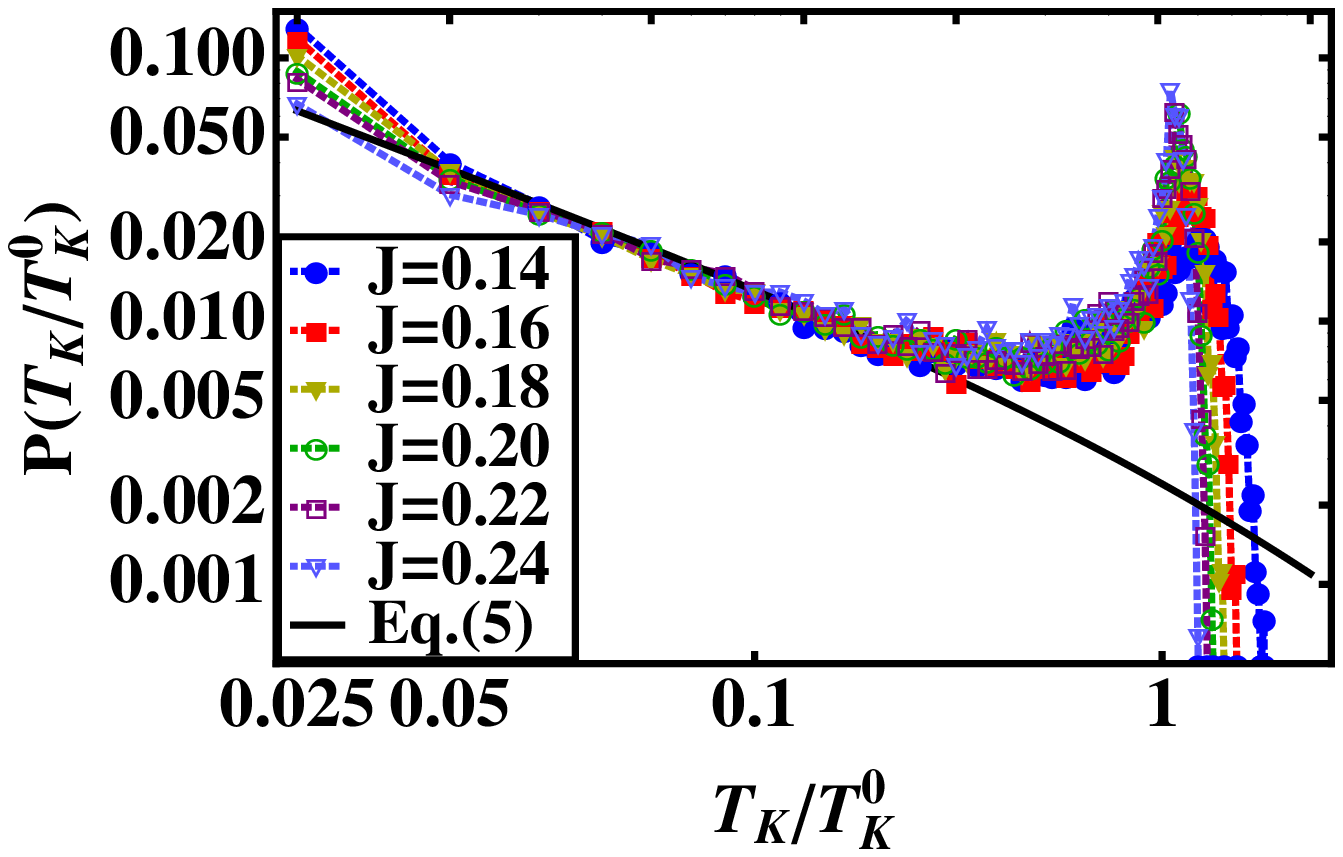}} 
 \caption{(Color online) $N(|J_{\rm RKKY}|) $ at (a) 
  fixed disorder strength $W=2t$, (b) fixed $R=5a$
  ($N=30\,000$, $L=100a$, $M=1000$). Black dashed lines: fit to
  log-normal distribution. (c) $P(T_K)$ 
  at fixed $j=J/D=0.25$, (d) $P(T_K)$ at fixed $W=5t$
  (N=30 000, L=40a, M=200). $E_F=2t$ in a)-d).} 
 \label{fig:rkky_disordered} 
\end{figure}

Eq.\,\eqref{eq:J-KPM} yields in a clean 2D system
%$
\begin{eqnarray}
  J^{0}_{\rm 2D} = - \frac{m^*}{8 \pi} \sin (2k_F R) /(k_F R)^2 \nn
\end{eqnarray}
in the asymptotic limit\,$k_F R \gg1$ with effective electron mass $m^*=1/(2a^2t)$ ,
and Fermi wave vector $k_F$. \cite{Kittel} 
%A direct comparison between the exact analytical expression
%in the clean limit and the numerical result yields excellent
%agreement. 
Its geometrical average is close to the clean limit for distances $R$ smaller
than  localisation length $\xi$, and decays exponentially at larger
distances,\cite{Sobota,HYLEE1}
$
 e^{{\la \frac{1}{2} \ln J_{\rm RKKY}}^2 \ra } \sim e^{-R/\xi}.
$
As shown in Figs.\,\ref{fig:rkky_disordered}\,a,\,b, the distribution of the absolute value
of $J_{\rm RKKY}$ is  well fitted by a log-normal, 
%$
\begin{eqnarray}
 N(x) = \frac{N}{\sqrt{2\pi \sigma}} \exp\left[
 -\frac{(x-x_0)^2}{2\sigma^2} \right], \nn
\end{eqnarray}
%$
where $x=\ln |J_{\rm RKKY}|$ and the fitting gives for $R= 5a$ and
$W=2t,4t$, $x_0=5,6$ and width $\sigma = 5.3 + .85 W/t$ increasing with the
disorder strength $W$. This is qualitatively consistent with analytical results obtained at weak
disorder,\cite{Lerner} while analytical calculations at strong disorder 
 have not been performed yet. This distribution width hardly depends on the distance
$R$. We used $N=30\,000$ disorder configurations. 

The distribution of $T_K$ is shown in Fig.\,\ref{fig:rkky_disordered}\,c, as
obtained from the numerical solution of Eq.\,\eqref{eq:nagaoka_kpm} for $L=40a$,
$j=J/D=0.25$. Since for every sample only one single site is taken to avoid a
distortion of the distribution due to intersite correlations, we had to use a
huge number of $N=30\,000$ different random disorder configurations to get
sufficient statistics. It has a strongly bimodal shape where the low $T_K$- peak
becomes more distinctive with larger disorder amplitude $W$.\cite{Grempel,
 Kettemann1, Kettemann4} In Fig.\,\ref{fig:rkky_disordered}\,d we show  these
results for fixed disorder strength $W=5t$ for various exchange couplings $j$.
Recently, an analytical derivation of the low $T_K$-tail of $P(T_K)$ was done,
using the multifractal distribution and correlations of
intensities.\cite{Kettemann5} These correlations are in 2D  logarithmic with
an amplitude of order $1/g$, where $g= E_F \tau$. For weak disorder, $g \gg 1$,
it corresponds to a power law correlation with power
$ 
\eta_{2D} = 2/ \pi g.
$
The correlation energy is of the order of the elastic scattering rate
$E_{c} \sim 1/\tau$. Thus, for $T_K \ll {\rm Max}\{\Delta_{\xi} = D/\xi^2,
 \Delta = D/L^2\}$,\cite{Kettemann5}
\begin{equation} \label{ptktailj}
 P(T_K) = \left( 1- p_{FM}
 \right)\left(\frac{E_c}{T_K} \right)^{1-j} ({\rm Min} \{ \xi, L \} )^{-\frac{d^2 j^2}{2\eta_{2D}}}, 
\end{equation}
where $p_{FM} = n_{FM}(0)/n = ({\rm Min} \{ \xi, L \} )^{-\frac{d^2 j^2}{2\eta_{2D}}}$, the ratio
of free PMs. Eq.\,\eqref{ptktailj} has a power law tail with power $\beta_j=1-j$
in good agreement with the numerical results,
Fig.\,\ref{fig:rkky_disordered}\,d, for $T_K/T_K^0 < .03$. For $T_K^0 > T_K >
{\rm Max}\{\Delta_{\xi} = D/\xi^2, \Delta = D/L^2\}$ one finds\cite{Kettemann5} 
\begin{eqnarray}\label{ptktail}
 \frac{P(T_K)}{ 1- p_{FM}}=
 (\frac{E_c}{T_K})^{1-\frac{\eta_{2D}}{2 d}}
 % \nn \\ \times
 \exp [ - \frac{(\frac{T_K}{E_c})^{\frac{\eta_{2D}}{d}}}{2 c_1} 
 \ln^2 \left(\frac{T_K}{T_K^0} \right) ], 
\end{eqnarray}
where $c_1 = 7.51$. This expression is in agreement with the numerical results,
see Fig.\,\ref{fig:rkky_disordered}\,d, using $\xi = g \exp (\pi g)$, and
$1/\tau = \pi W^2/6 D$, fitting only $E_c \approx .73 t$ and the
prefactor. Thus, we confirm that the power law tail  is governed by the
multifractal correlation with power $\eta_{2D}$. 

\begin{figure}[!Ht]
 \center
 \captionsetup[subfloat]{font = {bf,up}, position = top,
  captionskip=0pt, farskip=3pt} 
 \subfloat[] 
 {\includegraphics[width=0.25\textwidth]{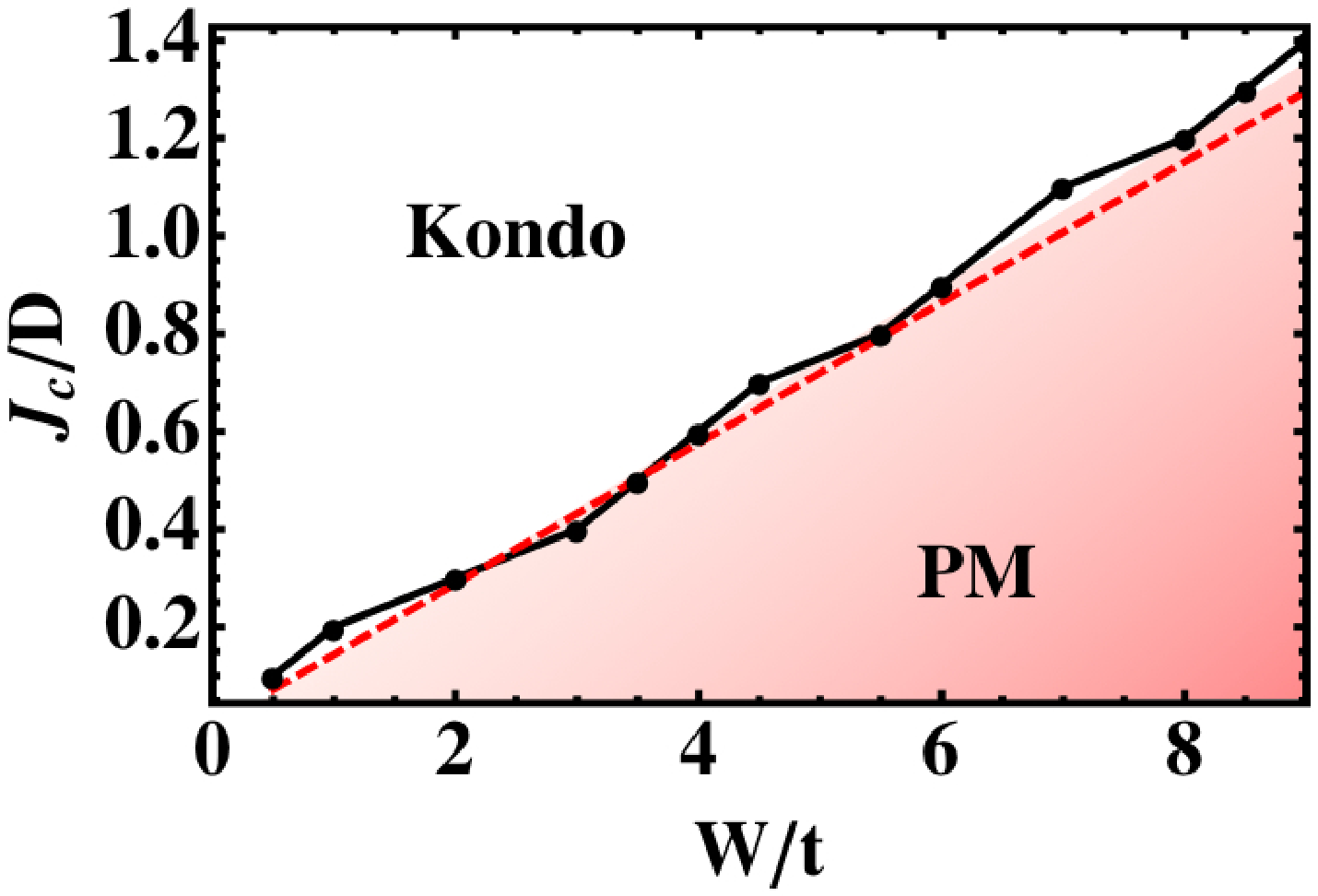}}
 \subfloat[]
 {\includegraphics[width=0.25\textwidth]{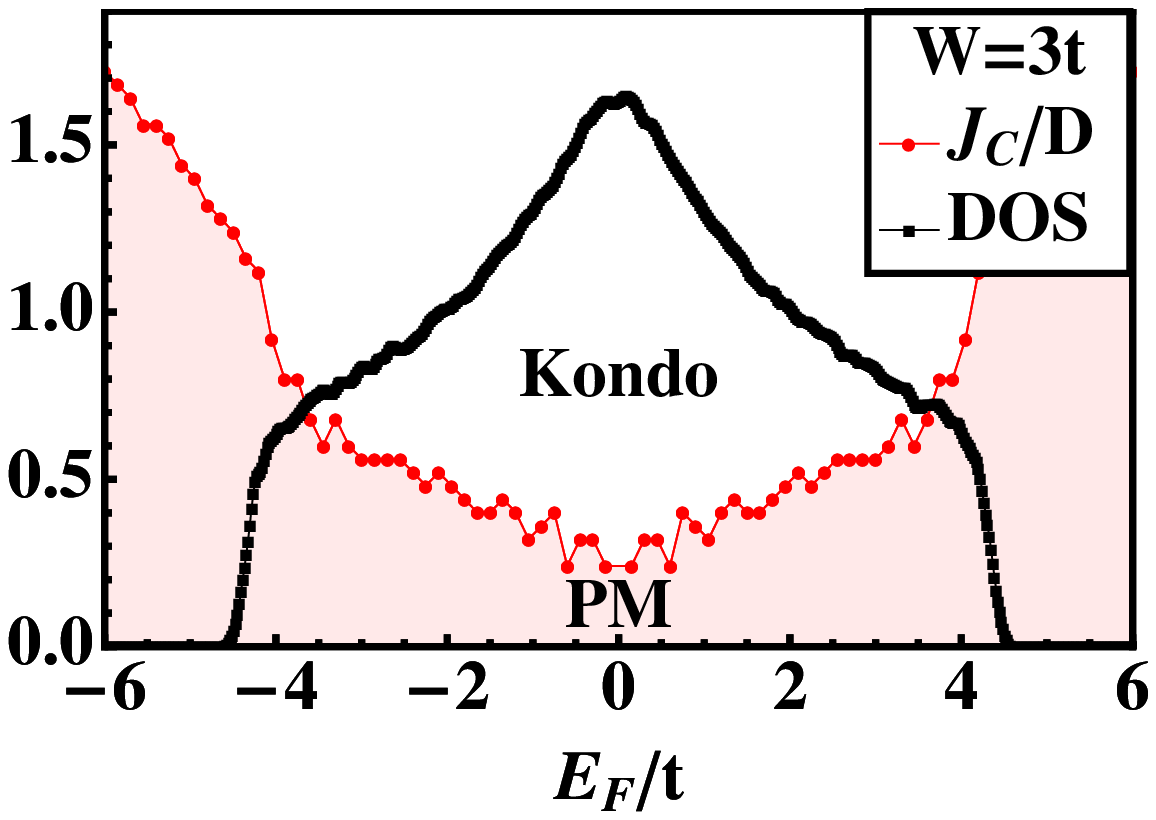}}
 \caption{(Color online) Quantum phase diagram with paramagnetic moment phase
  (PM) and Kondo screened phase: Critical exchange coupling\,$J_c/D$
  ($L=100$) (a) as function of $W$\,($M=300$,$E_F=2t$). Red dashed line:
  Eq.\,\eqref{eq:jc}. (b) as function of Fermi energy $E_F$ \,($M=200$); Red
  line: \,(DOS).} 
 \label{fig:jc_wrt_fermi}
\end{figure}

The quantum phase transition between the free paramagnetic moment phase (PM) and
a Kondo screened phase can be  studied by calculating the critical exchange
coupling $J_c$ above which there is no more than one free magnetic moment in the
sample volume $L^d$.\cite{Kettemann2} From the multifractality of the
eigenfunction intensities it is found to be related to the power $\eta_{2D}$ of
the power law correlations in the 2D DES as $J_{c} = \sqrt{ \eta_{2D}} D$ and
thus to increase in 2D linearly with disorder strength\,$W$ as,\cite{Kettemann5} 
\begin{equation}
  J_c = \sqrt{D/(3E_F)} W.
  \label{eq:jc}
\end{equation}
In Fig.\,\ref{fig:jc_wrt_fermi}\,a, \,Eq.\,\eqref{eq:jc} is plotted together
with numerical results as function of disorder strength\,$W$. We find good
agreement. There are only deviations at large disorder, $g <1$, where the $1/g$
expansion breaks down. 
%According to Eq.\,\eqref{eq:jc} $J_{c}$ increases as $E_F$ is moved towards the
%band edge. 
We plot $J_{c}$ as function of $E_F$ in Fig.\,\ref{fig:jc_wrt_fermi}\,b,
together with the density of states (DOS). We find that $J_{c}$ is increasing
towards the band edge as $1/\sqrt{E_F}$ in agreement with
Eq.\,\eqref{eq:jc}. Far outside of $\vep_{\rm edge}$ of the clean system it
increases as $J_c/D = 1/\ln | \vep_{\rm edge}- E_F|$  due to the gap
in the DOS. 
 
\section{Magnetic Phase Diagram at $T=0$}
In clean systems the critical density $n_c =1/R_c^d$ above which the MIs are coupled
with each other can be obtained from the condition that $|J^{0}_{\rm RKKY} (R_c) |
= T_K$.\cite{Doniach} Thus, in 2D with $|J^{0}_{\rm RKKY}|_{k_F R \gg 1} = J^2
\frac{m}{8 \pi^2 k_F^2 R^2}$ and $T_K = c E_F \exp (- D/J)$, $c \approx 1.14$,
 one finds
$ n_c = 16 \pi^2 c \frac{E_F^2}{J^2} \exp (-\frac{ D}{J})$.

In disordered systems, $T_K$ of an MI at a given site competes with the RKKY
coupling to another MI at distance $R$. Thus, the distribution function
$N(x_{JK})$ of the ratio of these two energy scales $x_{JK}=|J_{\rm RKKY}(R)| /
T_K$ for a given disordered sample with density of MIs $ n= 1/ R^2$, where $R$
is the average distance between the MIs, is crucial to determine  its magnetic
state. The distribution of $x_{JK}$ for $W=3t$ and $J/D=0.2$ is shown for
several distances $R$ in Fig.\,\ref{fig:ratio_kondo_rkky_square}\,(a)
($N=10\,000$, $L=100a$, $E_F= t$ and $M=300$). Likewise $N(T_K)$ and $N(J_{\rm
  RKKY})$, the distribution of $x_{JK}$ has an exponentially wide width
characterized by a small-$x_{JK}$ tails and a sharp upper cutoff in $x_{JK}$ as
shown in Fig.\,\ref{fig:ratio_kondo_rkky_square}\,(a). As increasing the
distance $R$ between the magnetic impurities the distribution $N(x_{JK})$ is
shifted to the left\,(smaller $x_{JK}$), since the RKKY interaction decreases
with $R$. 
The sharp upper cutoff in $x_{JK}$ allows us to define a critical density
$n_{c}(J)=1/R_c^2$ below which the Kondo effect dominates in the competition
with RKKY interaction at all sites. $n_{c}(J)$ is plotted in
Fig.\,\ref{fig:ratio_kondo_rkky_square}\,(b) for various values of disorder
strength $W$. When the MI density $n$ exceeds $n_c$, magnetic clusters start to
form at some sites and the MIs may be coupled by $J_{\rm RKKY}$. We see that
this coupled moment phase (CM) expands at the expense of the Kondo phase with
increasing $W$. When $R$ is larger than localisation length $\xi$ the coupling
$J_{\rm RKKY}$ is exponentially small and there is a {\it  paramagnetic phase}
(PM) below $n_{\xi} = 1/\xi(g)^2,$ where MIs remain free up to exponentially
small temperatures. 

\begin{figure}[!Ht]
 \center
 \captionsetup[subfloat]{font = {bf,up}, position = top,
  captionskip=0pt, farskip=0pt} 
 \includegraphics[width=0.4\textwidth]{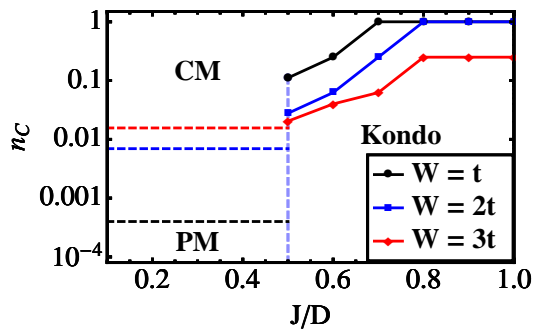}
 \caption{(Color online) Critical MI density $n_c$ as function of
  $J/D$ for graphene\,($\varepsilon_F=3t$, Dirac point) $L$, $M$, $N$
  as in Fig.\,\ref{fig:ratio_kondo_rkky_square}. }
 \label{fig:graphene_critical_n}
\end{figure}

In graphene the pseudogap at the Dirac point quenches the Kondo effect below
$J_c =D/2$, independently on disorder amplitude\,$W$. Thus, in graphene  there
is a larger parameter space where the MIs are coupled (CM) than in a normal
2DES, see Fig.\,\ref{fig:graphene_critical_n}. However, short range disorder
localises the electrons, cutting off the RKKY-interaction and for $n<n_{\xi}$
there is a PM phase. Thus, the magnetic phase in graphene is more stable against
Kondo screening but is more easily destroyed by disorder. 

\section{Doniach Phase Diagram of disordered 2DES and Graphene.}
\begin{figure}[!Ht]
  \includegraphics[width=0.5\textwidth]{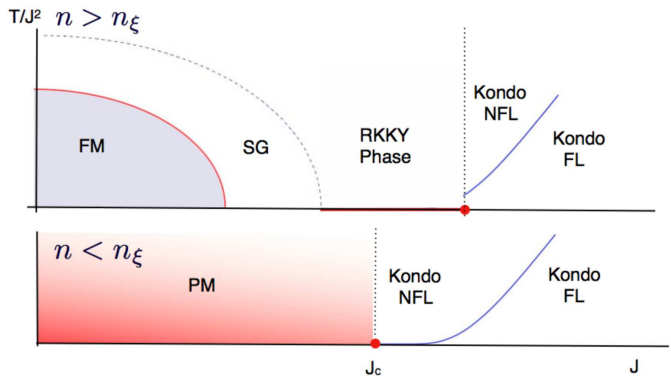}
 \caption{(Color online) 
  Schematic Doniach diagram: temperature $T$ divided by $J^2$ versus
  $J/D$. Vertical dotted line: critical point $J_c(n)$ separating RKKY phase
  from Kondo phase. Blue line: $T^*(n)$ separating Kondo FL phase from Kondo NFL 
  phase. For $n<n_{\xi}$ and $J<J_c$ a paramagnetic phase (PM) appears. }
 \label{fig:critical_n}
\end{figure}

We find, that the Kondo phase splits at finite temperature into a {\it Kondo
 Fermi-liquid (FL) phase}, where all MIs are screened, and a {\it Kondo
 Non-Fermi-liquid (NFL) phase}, at $T > T^*(n)$, where some MIs remain
unscreened and contribute to the magnetic susceptibility with an anomalous
temperature dependence, given by,\cite{Kettemann5}
% Thus, we find that the magnetic susceptibility 
% is diverging at low temperature in this Kondo NFL phase
%
\begin{eqnarray}
 \chi (T) \sim
 % \frac{n_{FM}(T)}{T} \sim n_{FM} (0)\frac{1}{T} +
 % \nonumber \\
 \frac{n}{E_{c}} 
 % \left\{ \begin{array}{cc}
 \frac{2d}{\eta_{2D}} \left(\frac{T}{E_{c}} \right)^{\frac{\eta_{2d}}{2d}-1}
 {\rm for\,} T >  T^*(n) > \frac{D}{\xi^2}. 
 % \frac{1}{j} \left(\frac{T}{E_{c}} \right)^{j-1} \xi^{-\frac{ 
 % 1}{2 \eta } (d j)^2}{\rm for} & T < \frac{D}{\xi^2}
 % \end{array}
 % \right..
\end{eqnarray}
%
% In $d=2$ it is a function of $W,$ $\eta_{2D} (W)= 2/ \pi g(W)$.
The temperature $T^*(n)$, plotted schematically in Fig.\,\ref{fig:critical_n}\, 
(blue line), is given by the position of the low $T_K$ peak in the
distribution $P(T_K)$, see Fig.\,\ref{fig:rkky_disordered}\,c. We note
that $J$ may be distributed itself and may add a nonuniversal,
material dependent contribution to the distribution of
$T_K$\cite{Wolfle} and $J_{\rm RKKY}$.  

For $n>n_c$ there is a succession of phases, starting with the {\it
  RKKY phase} where clusters are formed locally due to the widely
distributed RKKY coupling. Anomalous power laws are observed when
clusters are broken up successively as temperature is raised. From the
log-normal distribution $N(|J_{\rm RKKY}|)$ one obtains for the
magnetic susceptibility,
%$
\begin{eqnarray}
  \chi(T) T &=& n_{FM} (T) = \int_0^T d|J_{\rm RKKY}| N(|J_{\rm RKKY}|) \nn\\
  &\sim& n \exp \Big[ - \ln^2(T/|J_{\rm RKKY}^0|)/(2 \sigma(W))^2 \Big],
\end{eqnarray}
%$
where width $\sigma(W)$ increases with disorder strength $W$. Accordingly, the
excess specific heat is 
\begin{equation}
  C(T) = T \frac{d n_{FM}}{d T}\sim 
  \exp\Big[-\ln^2(T/|J_{\rm RKKY}^0|)/(2 \sigma(W)^2 \Big].
\end{equation}
The detailed analysis of the quantum phase diagram at  higher concentrations
$n$ requires  to go beyond our present analysis. One expects that at
$n > n_{SG}$ a {\it spin-glass phase} appears, where the magnetic susceptibility
shows a peak at spin glass temperature $T_{SG}$ as studied in
Refs.\,\onlinecite{Binder, Coqblin}. Above a critical density $n_{F}$ a phase with
long range order may form below a critical temperature
$T_c(n,J)$.\cite{Coqblin,Varma,Magalhaes,Bouzerar}
% It has been found earlier that $n_{SG}$ and $n_{F}$ do not strongly depend on 
% the exchange coupling $J$.\cite{Magalhaes}
%In graphene the pseudogap at the Dirac point quenches the Kondo effect below $J_c
%=D/2$, independently on disorder $W$. Thus, there is a larger parameter space where the
%MIs are coupled (CM), see Fig. \ref{fig:critical_n} (b). 
% However, short range disorder localises the electrons, cutting
%off the RKKY-interaction and for $n<n_{\xi}$ there is a PM phase.
% Thus, we conclude that in graphene the magnetic phase is  more stable against
% Kondo screening but is 
% more easily destroyed by disorder. 
%
%The recent discovery of topological insulators whose metallic surface states
% are protected by time reversal symmetry
% was followed by many experiments which study their stability to doping with 
% magnetic adatoms which can break the time reversal symmetry.\cite{ti}
%  Similarly attempts to make graphene magnetic by doping them with 
%  magnetic adatoms 

%

\section{Conclusions and Discussion}
We conclude that it is the full distribution function $N(x_{JK})$ of the ratio of the RKKY
coupling and the Kondo temperature  which determines the magnetic phase diagram
of magnetic moments in disordered electron systems, especially at low
concentrations. We  identified a critical density of magnetic impurities $n_c$
below which Kondo wins at all positions in a disordered sample above a critical
coupling $J_c$, which increases with the disorder amplitude. As a result, the
Kondo phase is diminished as the  disorder is increased, favoring  a phase
where the MI spins are coupled. The magnetic susceptibility obeys an anomalous
power law behavior, which crosses over as function of $J$ from the Kondo regime
where that power is determined by the wide distribution of the Kondo temperature
$T_K$,  to a  spin coupled phase  where it is governed by the log-normal
distribution of $J_{\rm RKKY}$. At low densities and small $J< J_c$, we identify
a paramagnetic phase. The distribution function of $|J_{\rm  RKKY}|/T_K$ is
expected to determine also the magnetic phase diagram of magnetically doped
graphene and the surface of topological insulators with magnetic adatoms, see
Fig. \ref{fig:graphene_critical_n}. This distribution function  may also be
crucial to explain the anomalous magnetic properties of doped semiconductors in
the vicinity of metal-insulator transition,\cite{Loehneysen} where we expect
that $\eta$ is replaced by the universal value $\eta = 2(\alpha_0-d)$, $d=3$
with the universal multifractality parameter $\alpha_0$.  

In this work  we  considered the distribution function of the Kondo temperature
of single impurities and the RKKY coupling of pairs of magnetic moments and
extracted information on the quantum phase diagram of systems with finite
concentrations of MIs. While this approach has its limitations, for example at
finite concentration  the RKKY coupling can reduce $T_K$ as has been already
found by Tsay and Klein in the 70s.\cite{Tsay1, Tsay2} However, they concluded
that this reduction is minor. More importantly, later work revealed that the
Kondo lattice of a finite density of magnetic moments, which is coupled to the
conduction electrons, has a coherent low temperature heavy fermion phase, and
a Kondo insulator phase at half filling of the magnetic moment levels. More
recently, the Kondo lattice  in 1 dimension was studied more rigorously (see the
review by Tsunetsugu et. al\cite{sigrist}), and it was shown  that, at least in
1D, the groundstate of this system can not be understood  by the mere extension
of the single and two-magnetic impurity problem, where the physics is governed
by the competition between these  two energy scales, the Kondo temperature and
the RKKY coupling. However, the higher temperature behavior was found to be
still governed by the competition between these two  energy scales. Therefore,
we expect that the consideration of the reduced problem of two impurity spins,
will give important information on the physics of disordered electron systems at 
finite concentration of magnetic moments, which becomes more meaningful the
lower the density and the higher the temperature is. Going beyond the
limitations of this approach, one will have to study the  disordered Kondo
lattice where a finite density of magnetic moments  is coupled to the conduction
electrons. For a clean Kondo lattice it is known that  a coherent low
temperature heavy fermion phase, and a Kondo insulator phase at half filling of
the magnetic moment levels appears.\cite{coleman,read} It remains to see how
these low temperature phases are modified by the presence of nonmagnetic
disorder.  

\acknowledgments
We gratefully acknowledge useful discussions with Georges Bouzerar, Ki-Seok Kim, 
Eduardo Mucciolo and Keith Slevin, as well as the support by the BK21
Plus funded by the Ministry of Education, Korea (10Z20130000023). 

%\section*{References}
\bibliographystyle{apsrev}
\bibliography{reference}

\end{document}